\documentclass[10pt, conference]{IEEEtran}
\usepackage[left=0.625in,right=0.625in,top=0.75in,bottom=0.99in]{geometry}

\IEEEoverridecommandlockouts
\usepackage{cite}
\usepackage{amsmath,amssymb,amsfonts}
\usepackage{graphicx}
\usepackage{color}
\usepackage{subfig}
\usepackage{stackengine}
\usepackage{comment}
\usepackage[normalem]{ulem}
\usepackage{caption}
\usepackage{tabularx}
\usepackage{bm}
\usepackage{dsfont}
\usepackage{mathtools}
\usepackage{lettrine}
\usepackage{comment}

\usepackage[
  draft,
  bookmarks=false
]{hyperref}

\usepackage[flushleft]{threeparttable} 
\usepackage{booktabs}

\usepackage{textcomp}
\usepackage{xcolor}

\usepackage[linesnumbered, ruled]{algorithm2e}
\SetKwRepeat{Do}{do}{while}%

\SetCommentSty{mycommfont}
\usepackage{algpseudocode}

\def\BibTeX{{\rm B\kern-.05em{\sc i\kern-.025em b}\kern-.08em
    T\kern-.1667em\lower.7ex\hbox{E}\kern-.125emX}}
    
\usepackage{orcidlink}

\usepackage{xspace}
\usepackage{ifthen}
\usepackage[inline]{enumitem}

\usepackage{algcompatible}

\newboolean{showcomments}
\setboolean{showcomments}{true}
\ifthenelse{\boolean{showcomments}}
{
}

\usepackage{subfig}
\usepackage{fancyhdr}

\begin{document}

\title{Semantic Edge-Cloud Communication for Real-Time Urban Traffic Digital Twins}

\author{Murat Arda Onsu$^1$, Arman Dogru$^1$, Poonam Lohan$^1$, Burak Kantarci$^1$, Aisha Syed$^2$, \\Matthew Andrews$^2$, and Sean Kennedy$^2$
\thanks{$^{1}$School of Electrical Engineering and Computer Science,
University of Ottawa, Ottawa, ON, Canada. 
Emails: \{monsu022, adogr056, ppoonam, burak.kantarci\}@uottawa.ca}
\thanks{$^2$Nokia Bell Labs, 600 March Road,
Kanata, ON K2K 2E6, Canada. \newline 
Emails: \{aisha.syed, matthew.andrews, sean.kennedy\}@nokia-bell-labs.com}
}

\maketitle
\begin{abstract}
Urban traffic systems are becoming increasingly complex, demanding real-time traffic perception, proactive collision prediction, and intelligent management to improve safety and reduce congestion. Although AI-driven traffic management has shown strong promise, its success relies on diverse, high-quality data spanning critical scenarios such as collisions, near-collisions, and traffic rule violations. Digital twins offer a powerful solution by enabling synchronized virtual replicas of physical traffic environments for real-time monitoring, collision prediction, traffic-flow optimization, and diverse realistic scenario generation. However, continuously synchronizing a high-fidelity urban traffic digital twin through edge surveillance cameras requires substantial edge-cloud communication, and transmitting raw video streams leads to severe bandwidth and latency constraints. To overcome this challenge, this paper proposes a semantic communication-based edge–cloud framework for real-time urban traffic digital twin applications. By leveraging edge intelligence to extract and transmit only task-relevant semantic traffic information instead of full video feeds, the proposed framework significantly reduces communication overhead while preserving the essential dynamics needed for accurate real-time digital twin reconstruction. Experimental results demonstrate more than 99\% transmission savings and an average trajectory reconstruction error of only 0.12 m, confirming the proposed framework as an effective, scalable, and low-latency solution for AI-enabled urban traffic digital twin applications.

\end{abstract}


\section{Introduction} \label{sec:1}
Rapid urbanization and increasing traffic density have made real-time traffic monitoring, collision prevention, and congestion mitigation essential in Intelligent Transportation Systems (ITS). Although AI has shown strong potential for urban traffic management, its performance depends on diverse and high-quality datasets that capture critical events such as collisions, near-collisions, off-lane behavior, and traffic rule violations. In practice, collecting such comprehensive datasets from real-world surveillance systems remains difficult, especially for rare yet safety-critical events.

Digital twin technology offers a promising solution by creating a synchronized virtual replica of a physical traffic environment. By reproducing roads, vehicles, traffic lights, pedestrians, and other dynamic elements in simulation, digital twins enable controlled scenario generation, large-scale data collection, and safe evaluation of intelligent traffic management strategies \cite{2}. Thus, they provide an effective link between real-world traffic systems and AI-driven decision-making. However, existing urban traffic digital twin studies often overlook the combined challenges of real-time synchronization, realistic environmental modeling, and communication (comm.) efficiency. In particular, continuously transmitting raw video streams from edge surveillance cameras to the cloud creates substantial bandwidth (BW) and latency overhead, limiting scalability and real-time operation. In addition, insufficient physical realism in vehicle behavior, actor placement, and trajectory consistency can reduce the reliability of the virtual representation.

To address these issues, this paper proposes a semantic comm.-enhanced edge-cloud digital twin framework for real-time urban traffic management. Building on our prior work on urban environment representation \cite{3}, the proposed framework reconstructs real traffic scenes on a cloud-hosted digital twin while transmitting only task-relevant semantic information extracted at the edge, rather than full camera feeds. Lightweight object detection modules integrated with surveillance cameras capture the essential properties of real-world actors, significantly reducing comm. overhead while preserving the dynamics needed for accurate reconstruction.

The proposed methodology consists of three stages. First, a lightweight object detection model deployed on edge surveillance cameras extracts bounding box coordinates of real-world traffic actors, such as vehicles. Second, these coordinates, together with vehicle tracking information, are semantically transmitted over the wireless edge-cloud link. Third, the cloud server reconstructs the real traffic scene in the digital twin by mapping image coordinates into simulation coordinates through a planar homography matrix derived from normal driving trajectory data \cite{1}. In addition, the virtual platform incorporates advanced physical modeling capabilities to represent vehicle motion, collision behavior, road interactions, and environmental responses more realistically.
The main contributions of this paper are summarized as follows:
\begin{itemize}
    \item We propose a semantic comm.-enhanced digital twin framework for constructing a realistic urban traffic environment on a virtual platform, enabling intelligent traffic management and safety-oriented urban mobility research.
    \item We integrate semantic comm. into the edge-cloud pipeline to transmit only task-relevant traffic information instead of raw video, achieving more than 99\% transmission savings and reducing comm. overhead and latency.
    \item We employ a planar homography-based mapping mechanism to transform real-world pixel coordinates into simulation coordinates, enabling accurate reconstruction of vehicle positions and activities in the digital twin with an average trajectory error of 0.12~m.
\end{itemize}

The rest of the paper is organized as follows. Section II reviews related work. Section III presents the proposed methodology, including dataset creation, edge-cloud deployment, semantic comm., and real-world digital twin reconstruction. Section IV reports the simulation setup and experimental results, and Section V concludes the paper.

\section{Related Works} \label{sec:2}

Digital Twin methodologies have been widely studied throughout the literature due to their capacity to create virtual representations of real-world systems, thereby enabling efficient monitoring, simulation, analysis, and optimization. A researcher in \cite{5} proposed the FlowTwin framework to address the challenges of validation costs and scalability issues in Digital twin development by introducing a dual-phase approach to MoDT development.  Another study in \cite{6} introduces a systematic, scalable methodology that employs a modular software architecture, incorporating an AI-driven sequential learning model to adapt dynamically to changing traffic conditions.  To minimize congestion and address the complexities of large urban areas, researchers in \cite{9} leverage digital twin technologies called DT-STOP that use real-time data from autonomous vehicles and roadside units. Moreover, in \cite{14}, edge computing and deep learning are jointly employed in digital twins to improve the real-time processing of traffic information and the performance of traffic sign recognition. Researchers in \cite{15} focus on optimizing urban mobility holistically, considering urban subsystems that can be influenced by factors such as weather, road hazards, traffic incidents, vehicle emissions, and so on. Therefore, they proposed a case study for Chattanooga in Tennessee with the Chattanooga Digital Twin, CTwin,  which enables situational awareness and cyber-physical control. However, these studies mainly overlook comm. latency and the minimum transmission payload during runtime, which limits the digital twin's adaptability to a real-time environment.

Semantic comm. is also studied across in digital twin domain. In \cite{11}, a semantic-driven traffic-generation framework for digital twin networks is proposed that combines semantic comm., Mininet-based simulation, and a transformer-based diffusion model to generate high-fidelity network traffic at lower comm. and computational costs. On the other hand, researchers in \cite{12} leverage a conditional generative semantic comm. framework methodology in a digital twin network to achieve reliable semantic recovery. Moreover, a point-cloud-based semantic comm. scheme for digital twins is introduced in \cite {13} to support efficient wireless transmission and accurate 3D scene generation under challenging channel conditions. Minimizing the synchronization latency of all user devices is studied in \cite{16} by introducing a soft actor-critic-based deep reinforcement learning algorithm in a semantic comm.-based digital twin.  Furthermore, a comprehensive framework for cognitive semantic comm. is developed by research \cite{17}, which optimizes key indicators such as semantic distortion, entropy, resource efficiency, latency, and security in digital twin platforms. Though these researches focus on the comm. part, they are not mainly considering the generation of the proper digital twin replica of the urban traffic.

In this work, we extend our previous study \cite{3} by introducing a real-time digital replica of the real world that is synchronized with low latency and BW efficient semantic edge-cloud comm. The proposed framework addresses the gap between computationally intensive high-fidelity traffic simulation and the severe BW limitations of edge-to-cloud networks, while enabling accurate reconstruction of a realistic urban traffic digital twin from real-world observations.

\begin{figure*}
  \centering
  \includegraphics[width=0.9\linewidth]{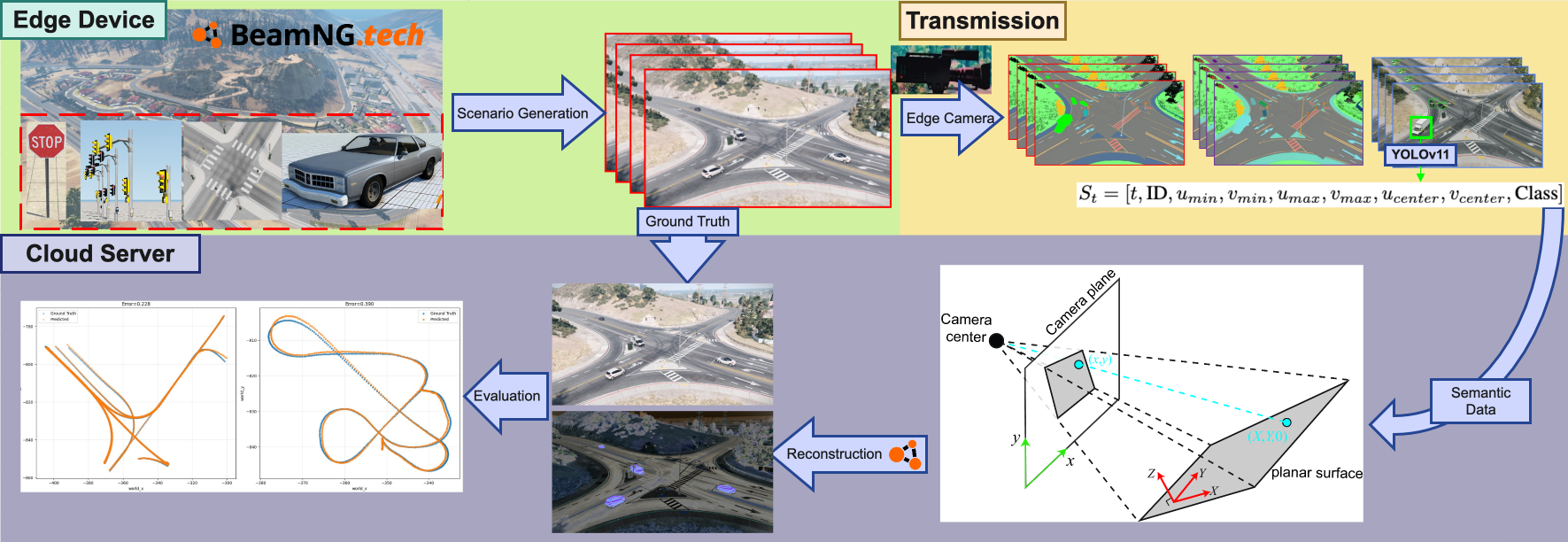}
  \caption{Proposed semantic communication-enhanced digital twin workflow in an edge-cloud architecture.}
  \label{fig:ontology}
\end{figure*}

\section{Methodology}



The proposed system architecture, which can be seen in \figurename \ref{fig:ontology}, is divided into three steps: (i)~edge devices, responsible for semantic extraction via computer vision models that distill each frame into a state vector, (ii) Edge-to-Cloud comm. to modulate and transmit the semantic data, which is the spatial and temporal information of the actors (eg. vehicles), and (iii)~a cloud server, responsible for spatial decoding and physics-based reconstruction within BeamNG.tech simulation platform~\cite{beamng_tech}. The proposed workflow offloads environment reconstruction to the cloud while using semantic comm. to  reduce BW use and latency compared to raw video transmission.


\subsection{Dataset Generation and Environment Modeling}

To ensure both flexibility and high visual fidelity, BeamNg.tech is adopted for digital twin environment modeling as a simulation tools for data generation and evaluation. Within this framework, BeamNG.tech reproduces urban and highway scenarios in a deterministic setting, enabling repeatable vehicle trajectories. 
A static virtual camera is then deployed to capture real-world actors (vehicles), under controlled conditions.

The experimentation dataset is generated by projecting the 3D world coordinates of procedurally generated traffic vehicles onto the camera sensor's 2D image plane. Let a vehicle's ground-truth 3D position be defined as $P_{world} = (X_w, Y_w, Z_w)$ and the camera's fixed spatial location as $C_{pos} = (C_x, C_y, C_z)$. The camera's orientation is defined by a quaternion $q = (q_x, q_y, q_z, q_w)$. Initially, the camera's orthonormal basis vector, Forward ($\vec{f}$), is derived from the quaternion:
\begin{equation}\small
\vec{f} = \begin{bmatrix} 2(q_x q_y + q_w q_z) \\ 1 - 2(q_x^2 + q_z^2) \\ 2(q_y q_z - q_w q_x) \end{bmatrix} 
\end{equation}
Assuming a global world vector $\vec{U}_w = (0,0,1)$, the Right ($\vec{r}$) and Up ($\vec{u}$) vectors are computed via cross products to ensure orthogonal projection mapping:
\begin{equation}\small
\vec{r} = {(\vec{f} \times \vec{U}_w)}/{\left\| \vec{f} \times \vec{U}_w \right\|}, \quad \vec{u} = {(\vec{r} \times \vec{f})}/{\left\| \vec{r} \times \vec{f} \right\|}
\end{equation}

The 3D world coordinate $P_{world}$ is then translated into camera's local relative coordinate, $P_{cam} = (X_c, Y_c, Z_c)$:

\begin{align}\small
X_c &= (P_{world} - C_{pos}) \cdot \vec{r} \\
Y_c &= (P_{world} - C_{pos}) \cdot \vec{u} \\
Z_c &= (P_{world} - C_{pos}) \cdot \vec{f}
\end{align}

Using the camera's vertical Field of View ($\text{FoV}_y$), aspect ratio $A = \frac{\text{Width}}{\text{Height}}$, and $Z_c$ as the depth scalar, the coordinates are projected into Normalized Device Coordinates (NDC) bounded by $[-1, 1]$:
{\small
\begin{equation}\small
\text{NDC}_x = \frac{X_c}{Z_c \cdot \tan\left({\text{FoV}_y}/{2}\right) \cdot A}, \; \text{NDC}_y = \frac{Y_c}{Z_c \cdot \tan\left({\text{FoV}_y}/{2}\right)} 
\end{equation}
}

Finally, these coordinates are scaled to absolute pixel coordinates $(u, v)$ corresponding to the video resolution $(W, H)$:
{\small
\begin{equation}\small
u = \left({\text{NDC}_x + 1}/{2} \right) \cdot W, \quad v = \left( 1 - {(\text{NDC}_y + 1)}/{2} \right) \cdot H 
\end{equation}
}

\subsection{Edge-Based Semantic Encoding of Traffic Images}
The proposed system converts pixel-level traffic images into a compact semantic payload by representing vehicles with bounding boxes instead of raw image data. After capture by edge cameras, the images are processed by a pre-trained YOLOv11 model deployed on edge hardware, such as surveillance camera. For each frame $t$, the edge node transmits only the encoded state vector $S_t$ of each detected vehicle, using bounding box coordinates, $(u_{box}, v_{box})$, thereby reducing redundant data and transmission overhead:
{\small
\begin{equation}
S_t = \left[ t, \text{ID}, u_{min}, v_{min}, u_{max}, v_{max}, u_{center}, v_{center}, \text{Class} \right] 
\end{equation}
}

\subsection{Spatial Decoding on the Cloud Server}

Upon receiving the encoded state vector $S_t$, the cloud server must decode them back into spatial world coordinates $(X_w, Y_w, Z_w)$. Because the framework assumes no vehicles are airborne, the predictive pipeline is strictly limited to Latitude and Longitude equivalents, with elevation derived deterministically from the simulation environment itself.

\subsubsection{2D Planar Homography}
Because vehicles are strictly ground-bound, the mapping between the camera's
image plane and the road surface can be described by a projective transformation
between two planes. This relationship is captured by a $3 \times 3$ planar
homography matrix $H \in \mathbb{R}^{3 \times 3}$, which encodes the full
perspective correspondence between pixel coordinates and world coordinates on
the ground plane.

A homography relates every point on one plane to a unique point on another
through a linear mapping in homogeneous coordinates. Given a pixel location
$(u, v)$ on the image plane, its homogeneous representation is the three-vector
$\tilde{\mathbf{p}} = [u,\; v,\; 1]^\top$. Applying $H$ yields a transformed
homogeneous vector:
{\small
\begin{equation}
\begin{bmatrix} x' \\ y' \\ w' \end{bmatrix}
= H \begin{bmatrix} u \\ v \\ 1 \end{bmatrix}
=
\begin{bmatrix}
h_{11} & h_{12} & h_{13} \\
h_{21} & h_{22} & h_{23} \\
h_{31} & h_{32} & h_{33}
\end{bmatrix}
\begin{bmatrix} u \\ v \\ 1 \end{bmatrix}
\end{equation}
}

The nine entries of $H$ encode three geometric operations simultaneously:
the upper-left $2 \times 2$ sub-matrix $(h_{11}$-$h_{22})$ captures rotation and
anisotropic scaling between the two planes; the elements $h_{13}$ and $h_{23}$
represent translation; and the bottom row $(h_{31},\; h_{32},\; h_{33})$ introduces
the projective component that accounts for foreshortening, i.e.,
the effect by which objects farther from the camera appear smaller in the image.
Because $H$ is defined up to an arbitrary scale factor, it has eight degrees of
freedom and therefore requires a minimum of four non-collinear point correspondences
to determine uniquely.

The output vector $[x',\; y',\; w']^\top$ is expressed in homogeneous coordinates,
where the third component $w'$ is a projective scaling factor that arises from
the perspective division. It does not correspond to a physical dimension; rather,
it encodes the depth-dependent scaling inherent in perspective projection. The
Euclidean world coordinates are recovered through dehomogenization:
{\small
\begin{equation}
\hat{X}_w ={x'}/{w'}, \quad \hat{Y}_w = {y'}/{w'}
\end{equation}
}

This division is the step that compensates for perspective distortion precisely because $w'$ varies across the image according to the depth
encoded by $H$. To estimate $H$, the Random Sample Consensus (RANSAC) algorithm is employed. RANSAC operates iteratively: in each iteration, four point correspondences are sampled at random from the traversable environment as pixel-world coordinate pairs, a candidate homography is computed from these four points via Direct Linear Transform (DLT), and the remaining correspondences are tested against this candidate. Points whose re-projection error falls below a predefined threshold of $\tau = 5.0$ pixels are classified as inliers. After a sufficient number of iterations, the candidate with the largest inlier consensus set is selected. At inference time, transforming a single pixel coordinate requires only one $3 \times 3$ matrix-vector multiplication and one scalar division, yielding a computational complexity of $\mathcal{O}(1)$ per vehicle per frame.

\subsubsection{Elevation Mapping}
While planar homography provides accurate 2D coordinate placements, real-world road topography features variations in elevation, such as hills and cambers, Operating under the strict kinematic assumption that passenger vehicles remain persistently grounded, spatial elevation ($Z_w$) is extracted deterministically using real-world Digital Elevation Model (DEM) data or the target environment's native terrain heightmap. The elevation is retrieved via a direct spatial lookup function $E$, which maps the decoded 2D coordinates to their corresponding topographical height:
{\small
\begin{equation}
\hat{Z}_w = E(\hat{X}_w, \hat{Y}_w) 
\end{equation}
}

\subsection{Digital Twin Reconstruction in BeamNG.tech}
The decoded $(\hat{X}_w, \hat{Y}_w, \hat{Z}_w)$ coordinates, combined with their temporal timestamps $t$, form a chronological trajectory matrix. This trajectory is shared to the cloud-hosted BeamNG.tech simulation. Utilizing the BeamNG's built-in AI Traffic Agents, the vehicles are initiated in their predicted locations.

To ensure similar dynamic behaviour, both vehicle orientations and velocities are calculated during reconstruction. First, the spatial displacement between sequential waypoints is computed. Let $\Delta t$ represent the elapsed time between state frames $t$ and $t-1$:
{\small
\begin{align}
\Delta X &= \hat{X}_{w, t} - \hat{X}_{w, t-1} \\
\Delta Y &= \hat{Y}_{w, t} - \hat{Y}_{w, t-1} \\
\Delta Z &= \hat{Z}_{w, t} - \hat{Z}_{w, t-1}
\end{align}
}
The scalar velocity (speed) $v_t$ of the vehicle is derived using the 2D Euclidean distance over the time interval:
{{\small
\begin{equation}
v_t = {\sqrt{\Delta X^2 + \Delta Y^2}}/{\Delta t}
\end{equation}
}

Similarly, the vehicle heading $\theta$ is derived so that the reconstructed vehicle is oriented along its direction of travel. Given the planar displacement components $\Delta X$ and $\Delta Y$ between two consecutive waypoints, the heading is computed as
{\small
\begin{equation}
\theta = \operatorname{arctan2}(\Delta Y,\Delta X)
\end{equation}
}
where $\operatorname{arctan2}(\cdot)$ returns the orientation angle of the displacement vector in the plane \cite{atan2}. This heading is then converted into a quaternion representation to set the physical orientation of the simulated vehicle:
{\small
\begin{equation}
q_z = \sin\!\left({\theta}/{2}\right), \quad q_w = \cos\!\left({\theta}/{2}\right)
\end{equation}
}

\subsection{ Edge-Cloud Communication}
\label{sec:system_model}
To evaluate the performance of Edge-to-Cloud communication, we model a 5G New Radio (NR) Frequency Range 1 (FR1) uplink channel. The system implements Low-Density Parity-Check (LDPC) coding and a Hybrid Automatic Repeat reQuest (HARQ) mechanism with Chase Combining \cite{morais20245g}.

\subsubsection{Physical Layer and Error Modeling}
The peak PHY layer data rate $R_{\text{PHY}}$ is a function of the allocated channel BW ($B = 100$ MHz), the number of bits per symbol for the chosen modulation scheme ($M \in \{\text{BPSK, QPSK, 16-QAM, 64-QAM}\}$), the LDPC coding rate ($R_c = 0.5$), and a fractional overhead $\eta$ accounting for guard bands and pilot signals. 

We model the wireless link as an Additive White Gaussian Noise (AWGN) channel. To accurately reflect 5G NR  channel performance, an empirical LDPC coding gain is applied to the base Signal-to-Noise Ratio (SNR). The theoretical uncoded Bit Error Rate (BER) is determined by the effective SNR and the specific modulation constellation. Data is segmented into fixed-size Transport Blocks (TB), and the Block Error Rate (BLER) is derived assuming independent bit errors within a block.

\subsubsection{HARQ with Chase Combining}
To ensure reliable payload delivery over the fading channel, our model implements a HARQ protocol utilizing Chase Combining (CC). In HARQ-CC, the receiver buffers the soft information of corrupted transport blocks and combines them with subsequent retransmissions. We model this energy accumulation as a linear effective SNR gain in the logarithmic domain ($G_{\text{CC}}$ in dB) for each retransmission round $k \in \{0, 1, \dots, K_{\text{max}}\}$. Consequently, the probability of a block failing decreases with each successive HARQ round.

\subsubsection{System-Level Latency and Goodput}
Difference between the transmission of raw images and YOLO CSV data yields vastly different payload sizes ($D$), defining system-level metrics accurately is critical. A payload is segmented into $N_{\text{blk}}$ transport blocks. During the $k$-th HARQ round, only the subset of pending (failed) blocks $N_{\text{pend}}^{(k)}$ is transmitted \cite{ahmed2021hybrid}. The total channel time elapsed is:

{\small
\begin{equation}\small
    T_{\text{total}} = \sum_{k=0}^{k_{\text{success}}} \left( \frac{D}{R_{\text{PHY}}} \cdot \frac{N_{\text{pend}}^{(k)}}{N_{\text{blk}}} \right) + k_{\text{success}} \cdot T_{\text{RTT}}
\end{equation}
}

where $T_{\text{RTT}}$ is the HARQ round-trip feedback delay.

If all blocks are successfully decoded within $K_{\text{max}}$ retransmissions, the payload Latency is recorded as $T_{\text{total}} + T_{\text{prop}}$, where $T_{\text{prop}}$ is the one-way propagation delay. If the HARQ limit is exhausted, the frame is dropped. To prevent artificially inflating the latency curves at low SNR, dropped frames are entirely excluded from the average latency statistics. 

Conversely, the System Goodput effective throughput is calculated across \textit{all} transmission attempts. For a successfully delivered payload, Goodput is $\Gamma = D / (T_{\text{total}}+T_{\text{prop}})$. Dropped frames contribute $0$ Mbps. This penalizes the average throughput at low SNR, correctly reflecting the system-level capacity degradation caused by large payload failures.

\subsection{End-to-End Computational Complexity Analysis}

The proposed architecture distributes the computational load asymmetrically. On the edge node, YOLOv11 inference is computed. Optimized for TensorRT or specific Neural Processing Units (NPUs), this operation scales with the input resolution $\mathcal{O}(W \times H)$ but requires extremely low memory BW. It entirely avoids the high energy costs and thermal throttling associated with deep hardware video encoding. 

On the receiver side, the cloud node executes $\mathcal{O}(1)$ matrix multiplications for the RANSAC homography decoding per vehicle, performs an $\mathcal{O}(1)$ spatial memory lookup for deterministic $Z$-elevation mapping via the DEM, and hosts the $\mathcal{O}(N)$ BeamNG.tech soft-body physics engine for $N$ vehicles. Consequently, the heavy, unbounded computational lifting is safely offloaded to centralized cloud environments where GPU/CPU resources, memory BW, and power consumption are unconstrained, leaving only lightweight extraction tasks at edge.

\section{Performance Evaluation}

In the framework setup, a YOLOv11 model with pretrained weights is used to detect vehicles at runtime. Simulation setup and experimental parameters are shown in Table \ref{tab:sim_params}.


\begin{table}[h]
\centering
\caption{Experimental Setup Parameters}
\label{tab:sim_params}
\begin{tabular}{lc}
\toprule
\textbf{Parameter} & \textbf{Value} \\
\midrule
Channel BW ($B$) & 100 MHz \\
Overhead Factor ($\eta$) & 0.76 \\
LDPC Coding Rate ($R_c$) & 1/2 \\
Transport Block Size ($N_{\text{TB}}$) & 8192 bits \\
Max HARQ Retransmissions ($K_{\text{max}}$) & 3 \\
HARQ Chase Combining Gain ($G_{\text{CC}}$) & 3.0 dB / round \\
HARQ RTT Delay ($T_{\text{RTT}}$) & 8.0 ms \\
Propagation Delay ($T_{\text{prop}}$) & 1.0 ms \\

Image Dimension ($WxH$) & 1920.0 x 1080.0 \\
Frame Per Seconds ($FPS$) & 30 \\
Total Number Of Samples & 8100 \\

\bottomrule
\end{tabular}
\end{table}

The experiment starts with a realistic representation of the real-world environment in the simulation platform using semantic data including spatial and temporal information, which include the detected vehicle coordinates and corresponding timestamps. 
Initially, the homography matrix, which maps pixel-wise coordinates collected via YOLOv11 to simulation coordinates, is generated from normal driving trajectories with multiple vehicles in dense traffic. Normal driving cases are performed in the 4-way junction map in the BeamNG simulation. To evaluate the homography matrix's performance, new vehicle trajectories collected in a non-ordered, random driving scenario with Unexplored locations are given to the matrix. According to the results, in a distance-wise comparison, the minimum distance between the ground truth digital coordinates and the reconstructed coordinates via the homography matrix is 0.0096 meters, the maximum is 0.4417 meters, and the average is 0.1260 meters. normal trajectories and evaluation trajectories can be seen in \figurename \ref{fig:res_1}.    

\begin{figure}
  \centering
  \includegraphics[width=0.85\linewidth]{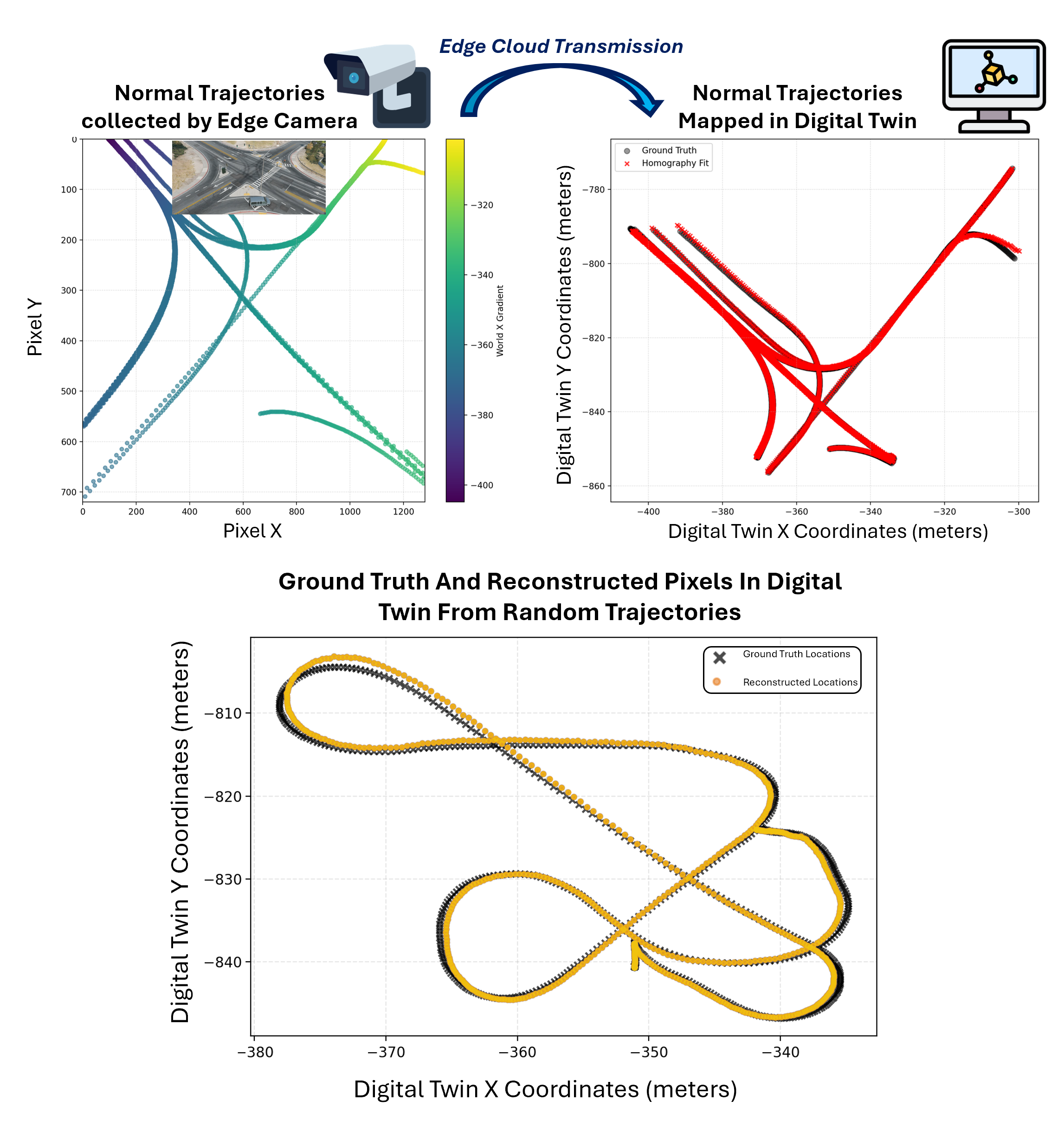}
  \caption{Real and Reconstructed Trajectories From Normal Driving (On The Top) and Random Driving (On The Bottom)}
  \label{fig:res_1}
\end{figure}

In the semantic communication part, edge-cloud data transmission is tested in the 5G new radio with 4 different modulation schemes called BPSK, QPSK, 16-QAM, and 64-QAM under various SNR values ranging from 1 to 30 dB. Evaluation is done by latency comparison between the transmission of raw image to pixel-wise coordinates and information of the objects in CSV format, and data rate. Results can be seen in \figurename \ref{fig:res_2}. 

\begin{figure}
  \centering
  \includegraphics[width=1\linewidth]{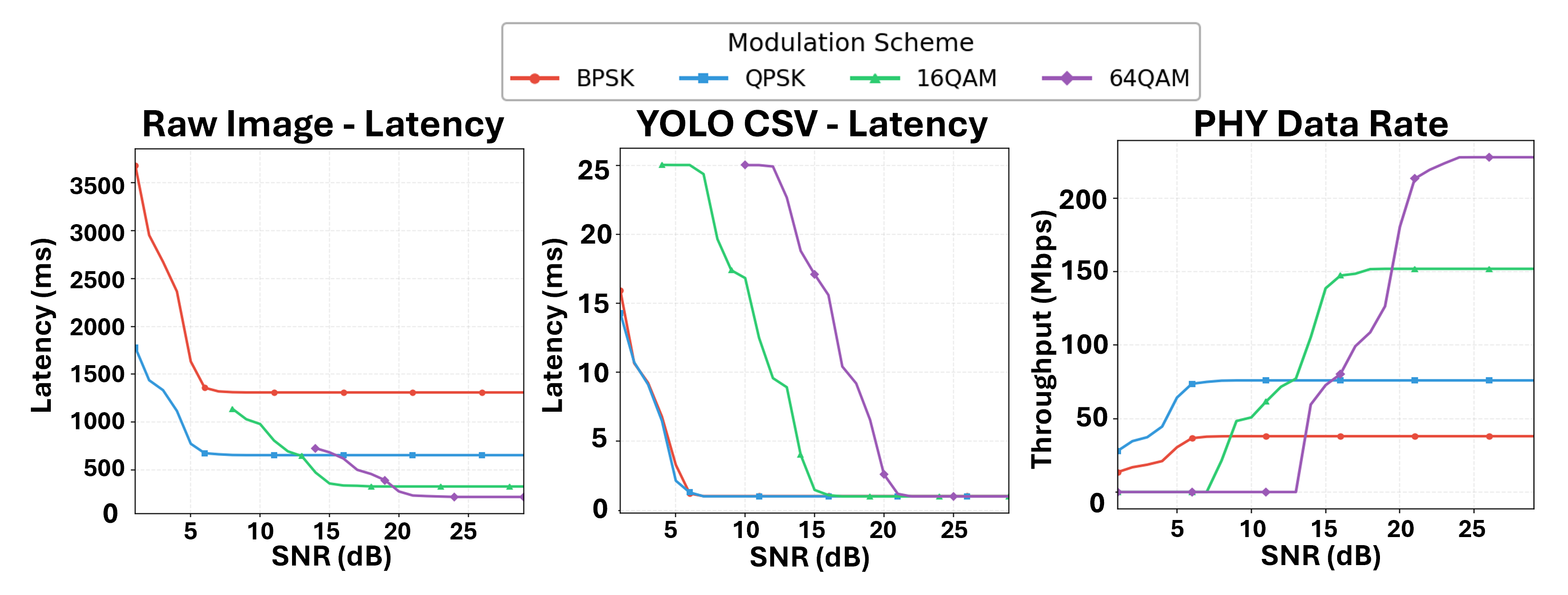}
  \caption{Latency and Data Rate Comparison of Raw Image and YOLO CSV Transmission under Different SNRs}
  \label{fig:res_2}
\end{figure}

According to \figurename \ref{fig:res_2}, for both image and CSV cases, the highest latency is achieved by BPSK, followed by QPSK, 16-QAM, and 64-QAM. On the other hand, the resilience to low SNR of low-bit modulation schemes such as BPSK and QPSK is much higher than of 16-QAM and 64-QAM. In the figure, if transmission fails for all packets, latency is considered to be 0, which illustrates that, in 16-QAM and 64-QAM modulation, the model can start transmission at 7 dB and 13 dB, respectively, for raw image data, while in the CSV case, these values are 4 dB and 10 dB. During the transmission of a raw image from edge to cloud, the range changes from 3500 to 500 ms, with the minimum achieved by 64-QAM at 21 dB SNR. On the other hand, the latency range for transmitting CSV data ranges from 25 to 1 ms. The real-time transmission, which is around 1 ms, is achieved in a 6 dB SNR ratio with BPSK and QPSK, while for 16-QAM and 64-QAM, these values are 16 dB and 21 dB, respectively. In terms of data Rate, Peak data rates for each modulation, BPSK, QPSK, 16-QAM, and 64-QAM, are 38, 76, 152, and 228, respectively. Although BPSK has the lowest peak data rate, its resilience it suitable for data transmission under low SNR values, that others cannot, making it suitable for low-adversarial networks. Results of transmission memory savings using YOLO detected coordinates transmission in CSV format is shown in \figurename \ref{fig:res_3}.

\begin{figure}
  \centering
  \includegraphics[width=0.95\linewidth]{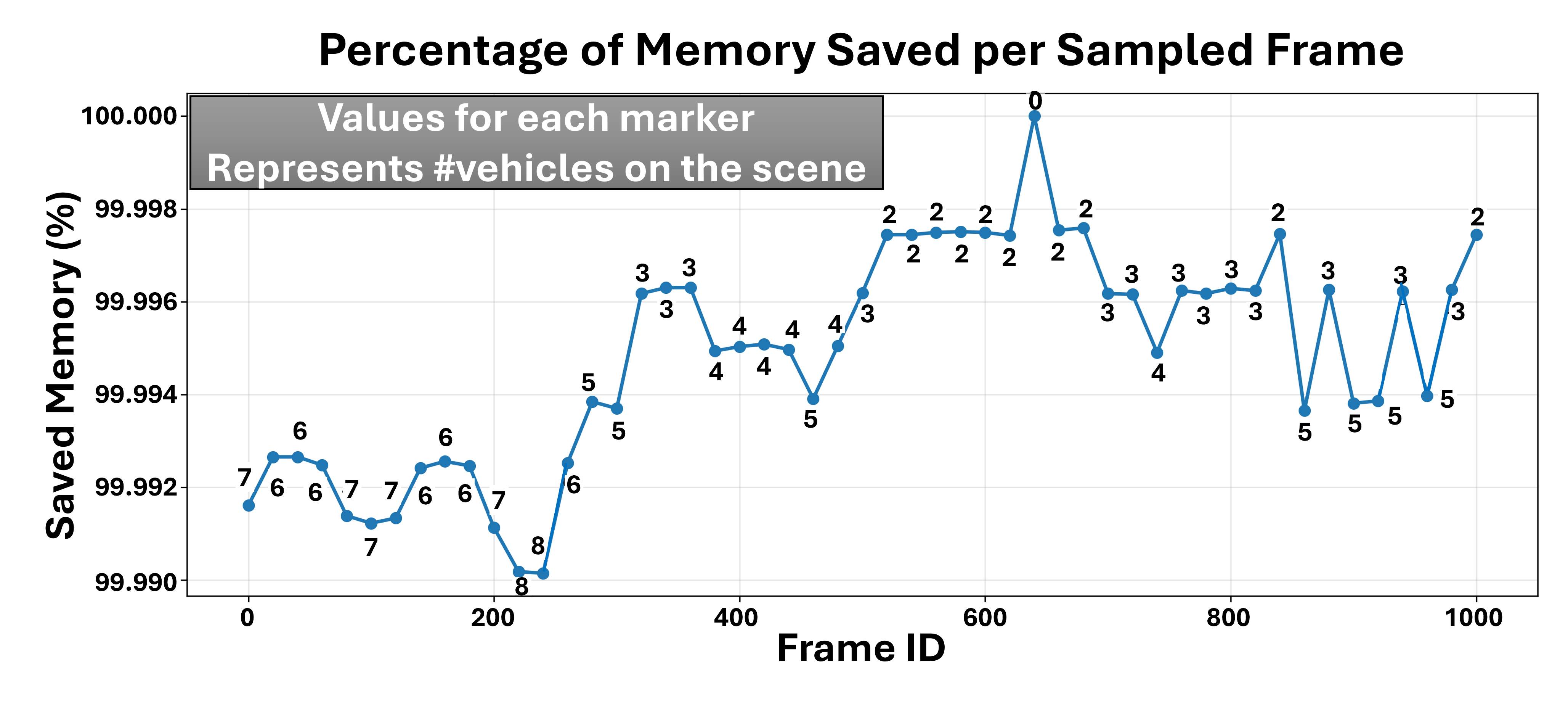}
  \caption{Memory Saving with CSV Transmission per Frame}
  \label{fig:res_3}
\end{figure}

\figurename \ref{fig:res_3} represents the memory saving, or in other words, the reduction in transmission data size in terms of percentages, with CSV format. The original image captured by the camera is 1920x1080 RGB, with a size of around 5.93 MB. On the other hand, each CSV row, which represents a single vehicle, consumes around 77.46 bytes, so the total memory consumption from 1 camera image is $N$ × 0.14 MB, where $N$ is the total number of cars detected by the YOLOv11 model. Across the overall runtime, the system achieved a memory saving within the range of 99.99\% to 100\%, where there is no visible vehicle on the scene. At runtime, the maximum visible number of vehicles on the frame is 8, which consumes around 774.60 bytes. Although the number of vehicles impacts the memory savings score, the linear increase in metadata payload remains statistically negligible relative to the large size of raw frames, ensuring that the data reduction efficiency consistently remains around 99.9\% even in high-density traffic scenarios.

\section{Conclusions} \label{sec:6}


In this paper, we proposed a semantic comm.-enhanced digital twin framework for real-time urban traffic management. By transmitting only task-relevant semantic information from edge surveillance cameras to the cloud, the proposed system significantly reduced edge-cloud communication overhead while preserving accurate and reliable synchronization between the real-world environment and its digital replica.
Experimental results in BeamNG.tech physics-based simulation platform validated the effectiveness of the proposed framework. The planar homography-based reconstruction achieved a mean error of only 0.1260~m, demonstrating accurate mapping of real-world observations into the simulation environment. In addition, the semantic comm. strategy achieved 99.99\% memory savings and reduced transmission latency from nearly 500~ms to approximately 1~ms under favorable SNR conditions. The framework also remained resilient in low-SNR settings, providing reliable synchronization even when conventional video streaming becomes ineffective.
Overall, these results show that the proposed framework is an effective, scalable, and low-latency solution for urban traffic digital twin applications. In future work, we will use this framework to generate diverse traffic scenarios, including collisions, traffic rule violations, and other safety-critical events, to support the development of advanced AI models for traffic flow optimization, collision avoidance, and related intelligent transportation tasks.

\section*{Acknowledgment }\label{Section6}
This work is supported in part by the MITACS Accelerate Program, in part by the NSERC CREATE TRAVERSAL program, and in part by the Ontario Research Fund-Research Excellence program under grant number ORF-RE012-026. 


\bibliographystyle{IEEEtran}
\bibliography{biblio.bib}{}


\end{document}